\documentclass[%
 reprint,
 amsmath,amssymb,
 aps,
]{revtex4-2}
\usepackage{mathptmx}
\usepackage{etoolbox}
\usepackage{xcolor}
\usepackage{hyperref}
\usepackage{graphicx}
\usepackage{dcolumn}
\usepackage{bm}

\begin{document}

\preprint{APS/123-QED}

\title{Noble-Gas Solubility in Solid and Fluid Metallic Hydrogen}

\author{Jakkapat Seeyangnok$^{1}$}
 \email{jakkapatjtp@gmail.com} 
\author{Udomsilp Pinsook$^{1}$}%
 \email{Udomsilp.P@Chula.ac.th}

\author{Graeme J Ackland$^{2}$}
 \email{gjackland@ed.ac.uk} 
\affiliation{$^{1}$Department of Physics, Faculty of Science, Chulalongkorn University, Bangkok, Thailand.\\
$^{2}$Centre for Science at Extreme Conditions, School of Physics and Astronomy, University of Edinburgh, Edinburgh, United Kingdom}%


\date{\today}

\begin{abstract}
Metallic hydrogen dominates the deep interiors of giant planets, where trace elements interact with dense quantum matter under extreme pressure. We investigate the thermodynamic stability of noble-gas impurities (He, Ne, Ar, Kr, Xe) in metallic hydrogen at 500 GPa using \textit{ab initio} molecular dynamics combined with first-principles free-energy calculations.  In the solid metallic phase, all noble gases exhibit positive formation free energies, driven by unfavorable electronic enthalpy and zero-point vibrational contributions. By contrast, heavier noble gases (Ar, Kr, Xe) appear soluble in liquid hydrogen, while He and Ne phase separate. This crossover reflects a competition between electronic repulsion and disorder-driven stabilization intrinsic to the liquid phase. Our results reveal noble-gas retention in metallic hydrogen, providing a microscopic mechanism for noble-gas fractionation in giant-planet interiors.
\end{abstract}

\maketitle

Hydrogen under extreme compression undergoes a  transformation from a molecular insulator to an atomic metallic state, a transition of central importance to both condensed matter physics and planetary science. This transition occurs in both solid and fluid phases, albeit at much lower pressure in the fluid\cite{knudson2015direct,silvera2010insulator,loubeyre2020synchrotron,Geng2019Thermodynamic,dzyabura2013evidence,van2021isotope,magduau2017theory,lue2024re,celliers2000shock,zaghoo2016evidence,scandolo2003liquid,morales2010evidence,mazzola2015distinct,attaccalite2008stable,holst2008thermophysical,morales2013nuclear,magdau2013identification,mcmahon2011ground,pickard2007structure,cheng2020evidence}.
In the deep interiors of giant planets such as Jupiter and Saturn, hydrogen is assumed to exist predominantly in a fluid metallic phase at pressures of several hundred gigapascals~\cite{guillot1999interiors,guillot2005interiors,militzer2008massive,mcmahon2012properties,miguel2023interior}. Under these conditions, electronic structure, bonding character, and phase stability differ fundamentally from ambient-pressure chemistry~\cite{hemley1998revealing,pickard2007structure,seeyangnok2025hydrogenation}, and even nominally noble-gas elements may display unexpected behavior~\cite{takezawa2025formation,liu2018reactivity,li2015stable,zhang2022pressure,zaleski2016krypton,zou2024exploring,peng2016unexpected,liu2024structural,takezawa2025formation}. Understanding impurity thermodynamics in metallic hydrogen is therefore essential for constructing microscopic models of planetary composition, differentiation, and atmospheric evolution.

The study of metallic hydrogen has long been driven by theory. Wigner and Huntington first predicted its existence within a free-electron framework~\cite{wigner1935possibility}, and subsequent first-principles investigations have mapped its structural evolution over wide pressure ranges~\cite{mcmahon2011ground,nagao1997structures,natoli1993crystal,barbee1991theoretical,ceperley1987ground,mcminis2015molecular,Geng2019Thermodynamic}. Near 500 GPa, atomic hydrogen is predicted to adopt the body-centered tetragonal $I4_{1}/amd$ phase, whose stability is governed by Fermi surface–Brillouin zone interactions rather than close packing~\cite{ackland2004origin}. Although experimental confirmation remains challenging and controversial~\cite{silvera2021phases,dias2017observation,eremets2011conductive,loubeyre2020synchrotron}, metallic hydrogen is widely predicted to host exotic properties, including high-$T_c$ superconductivity~\cite{ashcroft1968metallic,mcmahon2011high} and quantum fluid behavior~\cite{drummond2015quantum,monacelli2023quantum}.

At planetary conditions, hydrogen is unlikely to remain chemically pure but instead exists as a solution containing heavier elements~\cite{stevenson1982interiors}. Redistribution of dense core material into the surrounding metallic hydrogen envelope has long been proposed as a mechanism for core erosion in giant planets~\cite{stevenson1982interiors,chabrier2007heat,leconte2012new}. First-principles calculations have demonstrated that several candidate core constituents, including H$_2$O, MgO, SiO$_2$, Fe, and other alloys (Be, B, Mg, S, La), can dissolve in metallic hydrogen at extreme pressures and temperatures relevant to planetary interiors~\cite{wilson2012rocky,gonzalez2014ab,wahl2013solubility,seeyangnok2025solid}. These studies indicate that dissolution of heavy elements may be thermodynamically favored under metallic conditions, implying that dense materials are not confined to a distinct central core but are redistributed throughout the metallic hydrogen envelope, with a higher density towards the center due to the gravitational potential. Current models refer to "fuzzy cores" \cite{helled2024fuzzy} to describe such regions with no distinct boundary.

Among trace species, noble gases are particularly intriguing. Despite their closed-shell electronic configurations and chemical inertness at ambient conditions, they exhibit pronounced abundance anomalies in giant-planet atmospheres~\cite{guillot1999interiors,guillot2005interiors,wilson2010sequestration}. In particular, the depletion of neon in Jupiter’s atmosphere, as measured by the Galileo probe~\cite{mahaffy2000noble,owen1999low} provides compelling evidence for selective partitioning processes operating in the deep interior.  By contrast, excess argon and krypton suggest different behavior of the heavier Noble gasses. Elucidating the solubility and thermodynamic stability of noble gases in metallic hydrogen is therefore essential for understanding noble-gas fractionation, retention, and transport in planetary environments.

Previous theoretical efforts have concentrated primarily on hydrogen–helium immiscibility and phase separation~\cite{morales2009phase,wilson2010sequestration,militzer2013equation,lorenzen2009demixing} motivated by helium-rain scenarios in Jupiter and Saturn~\cite{stevenson1982interiors,fortney2010interior}. In parallel, first-principles studies have examined the dissolution of major core constituents in liquid metallic hydrogen~\cite{wahl2013solubility,wilson2012rocky,gonzalez2014ab}. However, systematic investigations of heavier noble gases (Ne, Ar, Kr, Xe) in metallic hydrogen remain scarce, particularly with explicit comparison between solid and liquid metallic phases. Because planetary interiors exhibit substantial thermal gradients, metallic hydrogen may exist in both crystalline and fluid states depending on depth and temperature~\cite{militzer2008massive,mcmahon2012properties}. Whether noble-gas solubility is intrinsically phase dependent under such extreme conditions therefore remains unresolved. Clarifying this issue requires a unified thermodynamic framework that treats noble-gas impurities consistently in both solid and liquid metallic hydrogen.

\section*{Methods}
    \begin{figure*}[ht]
    \centering
    \includegraphics[width=16cm]{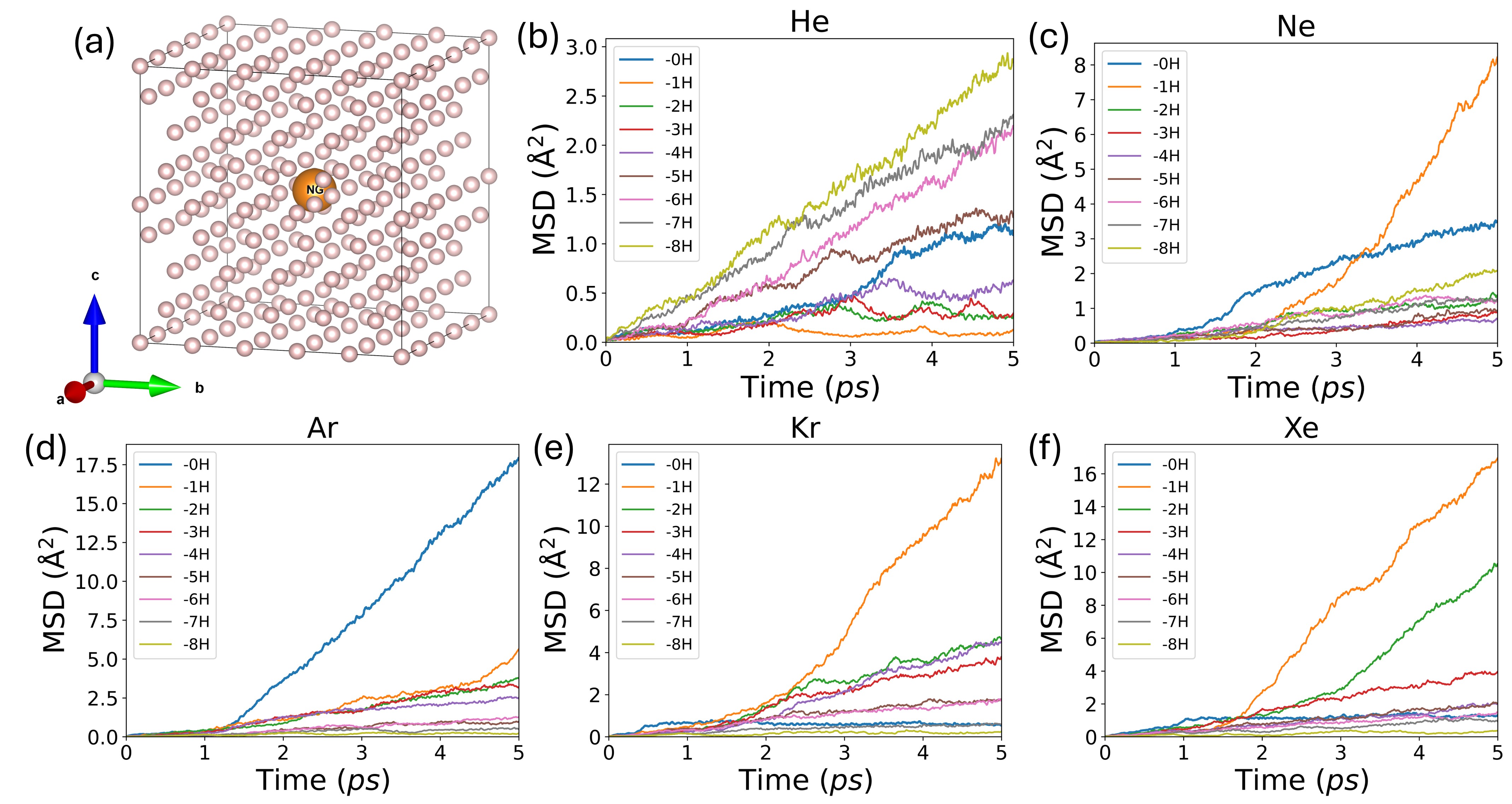}
    \caption{(a) Substitutional structure of a noble-gas impurity in solid metallic hydrogen. (b–f) Time evolution of the mean-square displacement (MSD) averaged over all atoms in alloyed solid metallic hydrogen for He, Ne, Ar, Kr, and Xe systems with different numbers of removed hydrogen atoms. For the configurations exhibiting the most stable MSD behavior—HeH$_{199}$, NeH$_{196}$, ArH$_{192}$, KrH$_{192}$, and XeH$_{192}$—the \textit{ab initio} molecular dynamics simulations were extended to 10.0~ps to ensure convergence of the enthalpy.}
    \label{fig:msd}
    \end{figure*}

Through relentless correct predictions, and despite obvious theoretical objections, density functional theory has established itself as the preeminent theory for electronic structure calculation\cite{jones2015density}.  Indeed it is often treated as ground-truth for machine-learned hydrogen potentials\cite{zong2020understanding,cheng2020evidence,istas2025liquid,kirsz2025tadah}.
    
\textit{Ab initio} molecular dynamics (AIMD) simulations were performed using \textsc{CASTEP}~\cite{CASTEP}. Solid metallic hydrogen was treated in the \textit{NPT} ensemble with a Berendsen thermostat and shape-changing Parrinello--Rahman barostat~\cite{ParRah} at $P = 500$~GPa and $T = 300$~K, allowing full cell relaxation. Liquid-phase simulations were carried out in the \textit{NPT} ensemble~\cite{NHC} with a constrained cubic cell using a Berendsen thermostat and an isotropic barostat (Andersen--Hoover) at $P = 500$~GPa and $T = 600$~K. A time step of 0.5~fs and the Verlet integration method \cite{VV} was used throughout. These AIMD simulations were performed using a $4\times4\times4$ $k$-mesh. The final AIMD configurations were fully relaxed within density functional theory (DFT) until the forces were smaller than $10^{-5}$~Ry/Bohr. Phonon properties for the solid phase were computed using density functional perturbation theory (DFPT)~\cite{baroni2001phonons} as implemented in \textsc{Quantum ESPRESSO}~\cite{giannozzi2009quantum}, using $8\times8\times8$ and $2\times2\times2$ $k$- and $q$-meshes, respectively, with wavefunction and charge density cutoffs of 60.0~Ry and 480.0~Ry, respectively, and a Methfessel--Paxton smearing of 0.04~Ry. The Perdew--Burke--Ernzerhof (PBE) generalized gradient approximation~\cite{perdew1996generalized} and Hartwigsen--Goedecker--Hutter norm-conserving pseudopotentials were employed. To investigate the bonding character, crystal orbital Hamilton population (COHP) analysis was performed using the LOBSTER package~\cite{deringer2011crystal,maintz2013analytic,maintz2016lobster}.

To analyze the results, we calculated the ensemble-averaged radial distribution function (RDF), 
\begin{equation}\label{RDF-equation}
g(r) = \frac{1}{4\pi r^2 N\rho} \sum_{i=1}^{N}\sum_{k\neq i}\left< \delta(r+r_{k}-r_{i})\right>
\end{equation}
and the  mean squared displacement (MSD)  defined as
\begin{equation}\label{MSD-equation}
    MSD(t) = \left< |\boldsymbol{x}(t)-\boldsymbol{x}(0)|^{2} \right>.
\end{equation}

The thermodynamic stability of a noble-gas impurity $X$ in hydrogen was quantified by the formation free energy
\begin{equation}
g = G(X\mathrm{H}_N) - G(X) - N H(\mathrm{H}),
\end{equation}
for a system containing $N$ hydrogen atoms and one impurity. Here $G(X\mathrm{H}_N)$ is the Gibbs free energies of the alloy supercell, and $H(\mathrm{H})$ and $G(X)$ are  the enthalpy per atom of hydrogen and the X reference impurity system respectively.

The Gibbs free energy was evaluated as
\begin{equation}
G = H + U_{\mathrm{ZPE}} - T(S_{\mathrm{vib}} + S_{\mathrm{conf}}),
\end{equation}
where $H$ was obtained from \textit{NPT} AIMD. The configurational entropy was treated within the ideal mixing approximation,
\begin{equation}
S_{\mathrm{conf}} = -k_B\left[c\ln c + (1-c)\ln(1-c)\right],
\end{equation}
where $c = \frac{N_{\mathrm{imp}}}{N_{\mathrm{imp}} + N_{\mathrm{H}}}$, and $N_{\mathrm{imp}}$ and $N_{\mathrm{H}}$ denote the numbers of impurity and hydrogen atoms, respectively.

Vibrational contributions were computed within the harmonic approximation from the phonon density of states $g(\omega)$. The zero-point energy and vibrational entropy were evaluated as
\begin{equation}
U_{\mathrm{ZPE}} = \int \frac{1}{2}\hbar\omega\, g(\omega)\, d\omega,
\end{equation}
\begin{eqnarray}
S_{\mathrm{vib}} &=& -k_B \int 
\ln\!\left[1-\exp\!\left(-\frac{\hbar\omega}{k_BT}\right)\right]
g(\omega)\, d\omega \nonumber \\
&+& \frac{1}{T}\int
\frac{\hbar\omega}{\exp(\hbar\omega/k_BT)-1}
g(\omega)\, d\omega .
\end{eqnarray}
The vibrational density of states was calculated either through density functional perturbation theory (DFPT) on the ideal lattice, of by fourier transform of the velocity autocorrelation function from MD.  In previous work, these have been demonstrated to be equivalent\cite{seeyangnok2025solid,seeyangnok2025hydrogenation}.

Reference structures at 500~GPa were $I4_1/amd$ hydrogen; hcp He, Kr, and Xe; fcc Ne; and dhcp Ar.

\section*{Results and Discussion}

\subsection*{Solid solubility}
    
    To evaluate the solid solubility, a combined molecular dynamics (MD) and static density functional theory (DFT) approach was employed. The enthalpy was obtained directly from \textit{NPT} ensemble MD simulations. By contrast, the phonon density of states (PhDOS) was calculated within the static DFT framework using DFPT based on atomic configurations relaxed from the final MD snapshots. 

    To capture the possibility that each impurity may substitute for more than one hydrogen, we performed \textit{ab initio} molecular dynamics simulations with progressively reduced hydrogen content. 
    In configurations with too many or too few hydrogens removed, we observed emission and diffusion of vacancy or interstitial defects from the impurity.
    
    This impurity  motion is reflected in the larger mean-square displacement (MSD) values shown in Figure~\ref{fig:msd}. The hydrogen-reduced system exhibiting a stable, approximately linear MSD therefore indicates that a dynamically equilibrated diffusive regime has been reached, while still preserving the crystal structure.  Such configurations were not the most stable and do not contribute to our solubility estimates.
    
    \begin{figure}[h!]
    \centering
    \includegraphics[width=8.5cm]{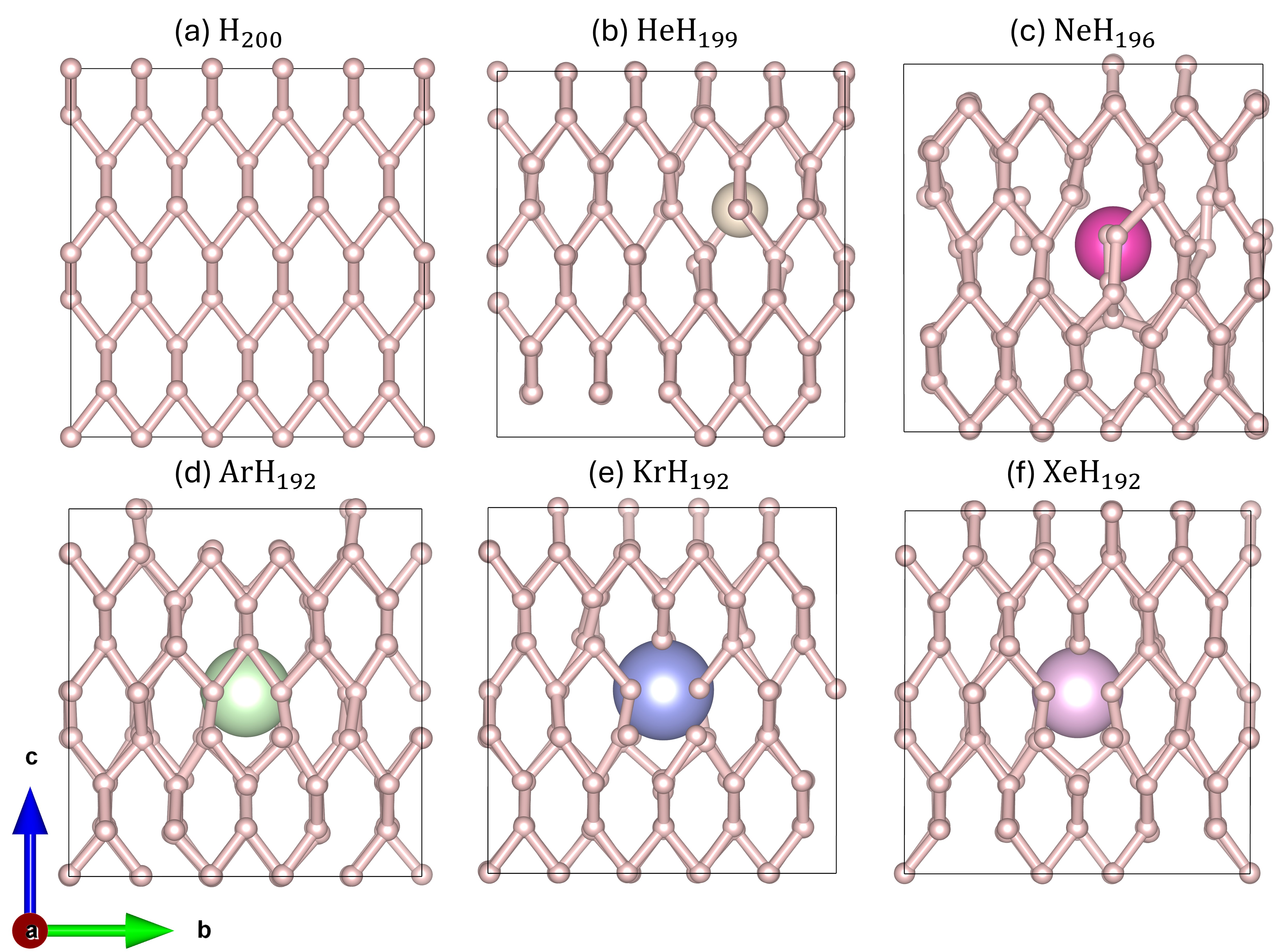}
    \caption{Optimized atomic configurations obtained from the final step of \textit{ab initio} molecular dynamics simulations at 500~GPa for (a) pristine metallic hydrogen (H$_{200}$) and noble-gas–substituted systems: (b) HeH$_{199}$, (c) NeH$_{196}$, (d) ArH$_{192}$, (e) KrH$_{192}$, and (f) XeH$_{192}$. Hydrogen atoms are shown in pink, while noble-gas impurities are highlighted in distinct colors.}
    \label{fig:optimized_MD}
    \end{figure}

\begin{figure}[h!]
    \centering
    \includegraphics[width=9cm]{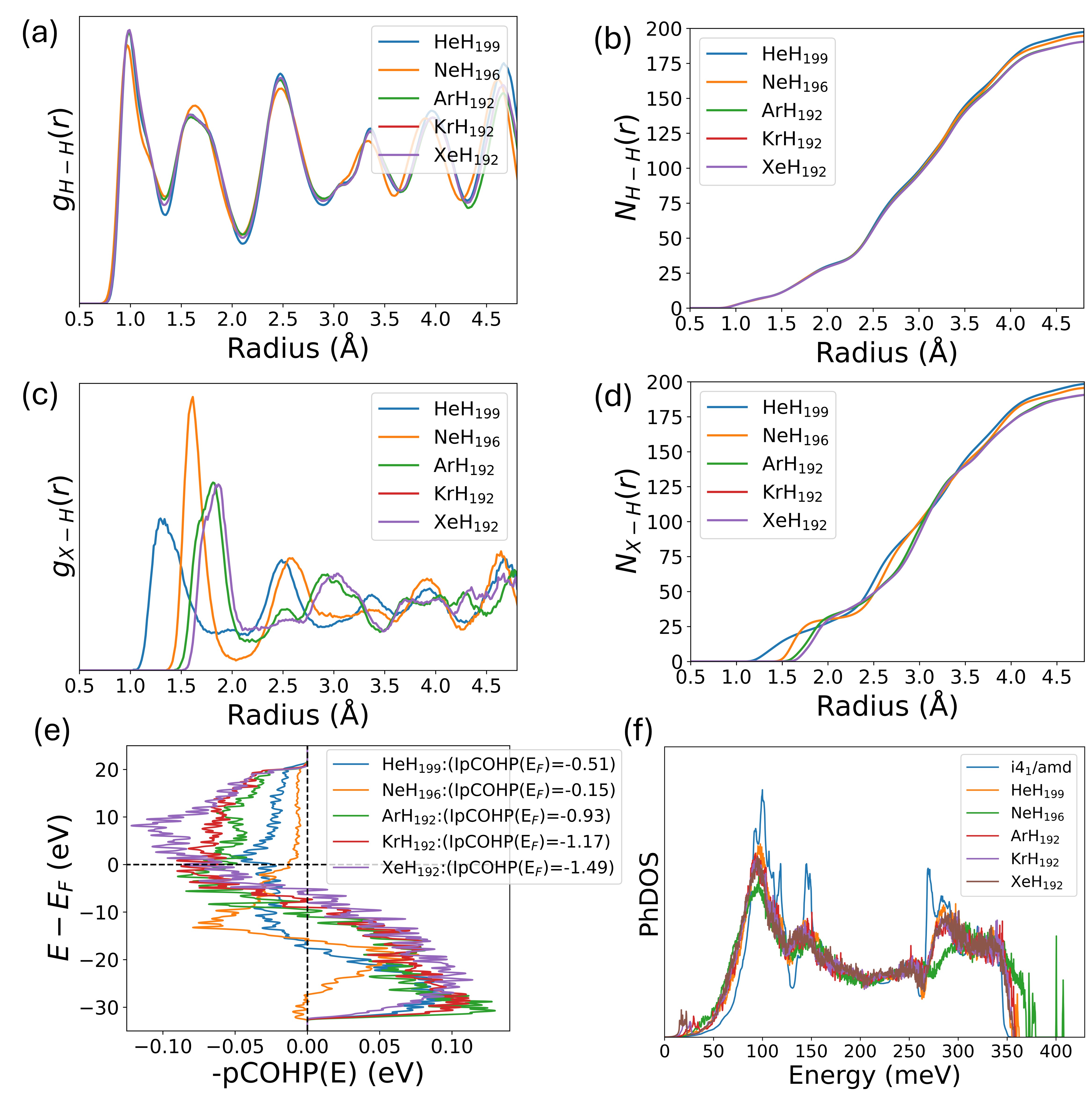}
    \caption{(a) Radial distribution functions (RDFs) of hydrogen in solid metallic $X\mathrm{H}_{Y}$ systems ($X = \mathrm{He}, \mathrm{Ne}, \mathrm{Ar}, \mathrm{Kr}, \mathrm{Xe}$). (b) Corresponding cumulative distribution functions (CDFs) describing the radial cumulative number of hydrogen atoms surrounding the noble-gas impurities. (c) RDFs of hydrogen around the impurity atoms $g_{X-H}(r)$. (d) Corresponding cumulative radial coordination numbers $N_{X-H}(r)$. (e) Crystal orbital Hamilton population (COHP) analysis for impurity--hydrogen interactions. (f) Phonon density of states (PhDOS) for pristine $I4_{1}/amd$ hydrogen and noble-gas-substituted systems.}
    \label{fig:phdos_rdf_cdf}
    \end{figure}

    The optimized configurations obtained from the final step of the AIMD simulations are illustrated in Figure~\ref{fig:optimized_MD}(a–f). The heavier noble gases occupy enlarged substitutional cavities within the hydrogen framework, leading to progressively stronger local lattice distortions with increasing atomic size. Despite these local distortions, the hydrogen framework largely preserves the characteristic $I4_{1}/amd$ network across all compositions, indicating that substitutional incorporation of noble gases does not trigger a global structural transformation. This structural robustness is further supported by the radial distribution functions (RDFs) between hydrogen atoms ($g_{H-H}(r)$) and noble-gas atom and hydrogen atoms ($g_{X-H}(r)$) shown in Figure~\ref{fig:phdos_rdf_cdf}(a) and (c), which retain the distinct peak structure characteristic of solid $I4_{1}/amd$ metallic hydrogen.

    The presence of enlarged substitutional cavities within the hydrogen framework is further reflected in the local hydrogen environment surrounding the noble-gas elements, as shown by the cumulative distribution functions (CDFs) of hydrogen ($N_{X-H}(r)$) in Figure~\ref{fig:phdos_rdf_cdf}(d). In contrast to the reconstructions of the first- and second-coordination shells reported previously for solid-solutions of more reactive atoms~\cite{seeyangnok2025solid}, the hydrogen positions are barely altered around the noble-gas impurities. The impurity simply replaces 1, 4 or 8 hydrogens for He, Ne and heavier elements respectively (Fig.\ref{fig:optimized_MD}). The absence of reconstruction indicates weak chemical interaction between the noble gases and the surrounding hydrogen network. 
    
    This weak interaction is further confirmed by the crystal orbital Hamilton population (COHP) analysis between the noble-gas impurity and the closest hydrogen atom, as shown in Figure~\ref{fig:phdos_rdf_cdf}(e). The small integrated projected COHP (IpCOHP) values at the Fermi level ($E_F$) indicate negligible covalent bonding between noble-gas atoms and hydrogen, with IpCOHP values of $-0.51$, $-0.15$, $-0.93$, $-1.17$, and $-1.49$~eV for He, Ne, Ar, Kr, and Xe, respectively. These values are substantially smaller in magnitude than those reported for hydrogen-rich systems such as saturated CH$_6$, where strong covalent bonding interactions give rise to an average IpCOHP of approximately $-7.15$~eV~\cite{seeyangnok2025hydrogenation}.
    
    Consistent with this picture, the phonon density of states (PDOS) as shown in Figure~\ref{fig:phdos_rdf_cdf}(d) reveals no additional high-frequency hydrogen vibrational modes that would indicate the formation of covalent impurity–hydrogen bonds. Instead, the noble-gas atoms introduce localized low-energy vibrational features, characteristic of mass-induced changed in acoustic modes. The high-frequency region of the spectrum remains dominated by hydrogen vibrations, indicating that the intrinsic hydrogen bonding network is largely preserved upon impurity incorporation, with only a minor contribution at the high-energy edge. In contrast to hydrogen-rich systems such as saturated CH$_6$, where distinct high-frequency modes originate from covalent bonding interactions~\cite{seeyangnok2025hydrogenation}, the vibrational signatures observed here arise primarily from the large atomic mass and size of the noble-gas impurities at low frequency, and a stiffening of the highest frequency modes as shown in Figure~\ref{fig:phdos_rdf_cdf}(f).

    Neon is an exception to this picture - it generates two distinct H$_2$ stretching modes around 400~meV - weaker than the 516~meV of molecular H$_2$ but significantly higher than the metallic modes.  These induced molecules lead to a high zero-point energy contribution from the coordination around Ne as shown in Table~\ref{tab:enthalpy_XH192}.

    \begin{table}[h!]
    \centering
    \caption{Thermodynamic contributions to the formation free energy $g$ of noble-gas impurities in metallic hydrogen. The enthalpy difference $\Delta H$, zero-point energy contribution $\Delta U_{\mathrm{ZPE}}$, entropic contribution $T\Delta S$, and the resulting formation free energy $g$ are shown, with all quantities reported per simulation cell in the unit of eV/cell.}
    \label{tab:thermo_ng_metallic_H}
    \begin{tabular}{lcccc}
    \hline\hline
    System & $\Delta H$ & $\Delta U_{\mathrm{ZPE}}$ & $T\Delta S$ & $g$ \\
    \hline
    HeH$_{199}$ & $2.70 \pm 0.07$ & $3.68$ & $0.09$ & $6.29 \pm 0.07$ \\
    NeH$_{196}$ & $4.69 \pm 0.07$ & $5.12$ & $0.17$ & $9.64 \pm 0.07$ \\
    ArH$_{192}$ & $2.36 \pm 0.06$ & $3.64$ & $0.13$ & $5.87 \pm 0.06$ \\
    KrH$_{192}$ & $1.94 \pm 0.08$ & $3.34$ & $0.14$ & $5.14 \pm 0.08$ \\
    XeH$_{192}$ & $1.25 \pm 0.07$ & $3.20$ & $0.16$ & $4.30 \pm 0.07$ \\
    \hline\hline
    \end{tabular}
    \end{table}

    The calculated thermodynamic decomposition reveals that noble-gas impurities are not thermodynamically stable in metallic hydrogen. As summarized in Table~\ref{tab:thermo_ng_metallic_H}, all systems exhibit a large positive enthalpic contribution $\Delta H$, originating from the electronic part of the energy. This indicates that the incorporation of a closed-shell noble-gas atom into the metallic hydrogen matrix is energetically unfavorable, primarily due to strong Pauli repulsion and the absence of chemical bonding between the impurity and the surrounding hydrogen atoms. The magnitude of $\Delta H$ dominates the free-energy balance and constitutes the primary driving force against dissolution.

    In addition to the electronic enthalpy penalty, the zero-point energy contribution $\Delta U_{\mathrm{ZPE}}$ is also positive for all noble gases considered. This reflects the stiffening of local vibrational modes induced by the presence of the impurity, which further increases the total free energy. Although the entropic contribution $T\Delta S$, arising from both vibrational and configurational entropy, is slightly favorable, its magnitude remains too small to compensate for the combined enthalpic and zero-point energy penalties. As a result, the total formation free energy $g$ remains large and positive in all cases, demonstrating that noble gases do not form solid solutions in metallic hydrogen under the conditions studied.

\subsection*{Liquid solubility}
\begin{figure}[h!]
    \centering
    \includegraphics[width=9cm]{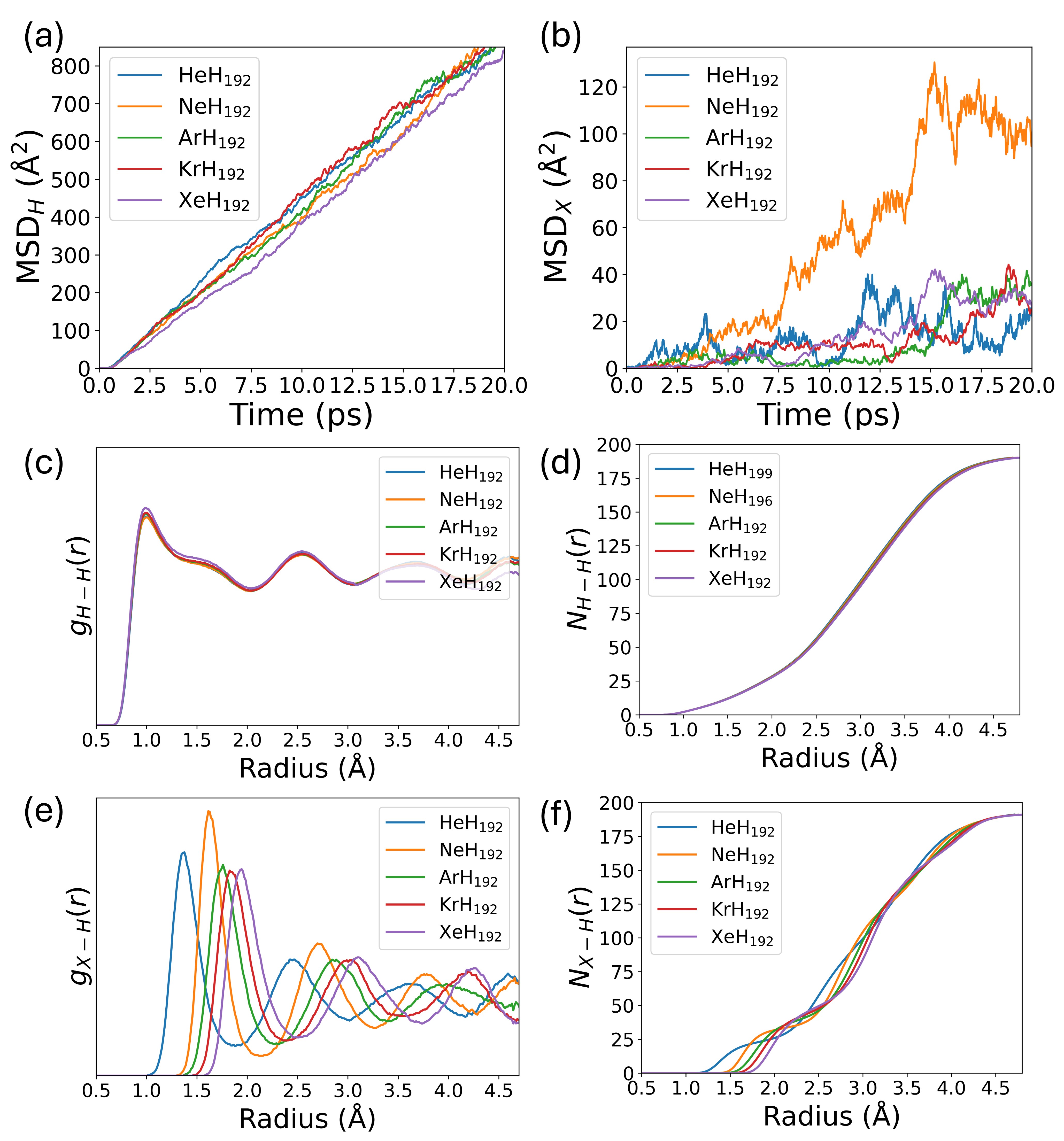}
    \caption{Mean-square displacement (MSD) and structural correlations in noble-gas–alloyed liquid metallic hydrogen at 500 GPa and 600 K within the $NPT$ ensemble. (a) Time evolution of the hydrogen MSD, $\mathrm{MSD}_H$, for HeH$_{192}$, NeH$_{192}$, ArH$_{192}$, KrH$_{192}$, and XeH$_{192}$. (b) Time evolution of the noble-gas impurity MSD, $\mathrm{MSD}_X$. (c) Radial distribution functions of hydrogen–hydrogen pairs, $g_{H-H}(r)$. (d) Corresponding cumulative coordination numbers $N_{H-H}(r)$. (e) Radial distribution functions of impurity–hydrogen pairs, $g_{X-H}(r)$. (f) Corresponding cumulative coordination numbers $N_{X-H}(r)$.}
    \label{fig:msd_rdf_liquid}
    \end{figure}
    
    As shown in Figure~\ref{fig:msd_rdf_liquid}(a), the approximately linear increase of the hydrogen mean-square displacement (MSD) with time indicates diffusive liquid behavior under these extreme conditions. In contrast, the MSD slopes of the noble-gas atoms in Figure~\ref{fig:msd_rdf_liquid}(b) are significantly smaller than those of hydrogen, reflecting the much slower motion of the heavier impurity atoms in the hydrogen matrix. Although the hydrogen MSD curves remain similar for different impurities, slight variations are observed, with heavier noble gases exhibiting marginally reduced mobility. This behavior suggests that larger and heavier impurities introduce local structural perturbations that mildly hinder hydrogen diffusion without significantly altering the overall liquid dynamics.

    The RDFs of $g_{H-H}(r)$ as shown in Figure~\ref{fig:msd_rdf_liquid}(c) exhibit a pronounced first coordination peak near $\sim$1~\AA, characteristic of dense atomic hydrogen, followed by a shoulder at slightly larger distances and damped oscillations indicative of short-range order without long-range crystallinity. The rapid decay of these oscillations beyond the second and third coordination shells confirms the absence of structural ordering typical of a solid phase, as further supported by the smooth behavior of $N_{H-H}(r)$ shown in Figure~\ref{fig:msd_rdf_liquid}(d). Importantly, the overall similarity of the RDF profiles demonstrates that noble-gas incorporation does not significantly disrupt the underlying hydrogen framework. The position of the first peak remains nearly unchanged across compositions, indicating that the average H–H separation is preserved even in the presence of impurities. Subtle differences in peak height and broadening reflect size-dependent local structural effects, as suggested by $g_{X-H}(r)$ and $N_{X-H}(r)$ shown in Figure~\ref{fig:msd_rdf_liquid}(e) and (f), with heavier noble gases inducing slightly enhanced local structuring. Together, the MSD and RDF analyses confirm that all systems remain homogeneous liquids and that noble-gas solubility in metallic hydrogen does not arise from structural phase transitions, but from thermodynamic stabilization mechanisms.
    
    To quantify thermodynamic stability in the liquid phase, we evaluate the formation enthalpy, $dH$, defined as
    \begin{equation}
    dH = H(X\mathrm{H}_{192}) - H(X) - 192\,H(\mathrm{H}),
    \end{equation}
    where $H(X\mathrm{H}_{192})$, $H(X)$, and $H(\mathrm{H})$ denote the enthalpies of the alloyed system, the elemental impurity, and hydrogen, respectively. A negative value of $dH$ indicates thermodynamic stability relative to the separated constituents.

    \begin{table}[h!]
    \centering
    \caption{Enthalpy components and formation enthalpy $dH$ for liquid metallic $X$H$_{192}$ systems ($X = \mathrm{He}, \mathrm{Ne}, \mathrm{Ar}, \mathrm{Kr}, \mathrm{Xe}$) obtained from \textit{NPT} simulations at $T = 600$~K and $P = 500$~GPa. The quantities $H(X\mathrm{H}{192})$, $H(X)$, and $H(\mathrm{H})$ are given in units of eV/atom, while $dH$ is reported in eV/cell.}
    \label{tab:enthalpy_XH192}
    \begin{tabular}{lcccc}
    \hline\hline
    X &
    $H(X\mathrm{H}_{192})$ &
    $H(X)$ &
    $H(\mathrm{H})$ &
    $dH$ \\
    \hline
    He & $-9.94 \pm 0.11$ & $-69.14 \pm 0.00$ & $-9.63 \pm 0.11$ & $0.14 \pm 0.15$ \\
    Ne & $-14.43 \pm 0.07$ & $-936.45 \pm 0.00$ & $-9.63 \pm 0.11$ & $0.53 \pm 0.13$ \\
    Ar & $-12.52 \pm 0.08$ & $-565.53 \pm 0.00$ & $-9.63 \pm 0.11$ & $-1.51 \pm 0.13$ \\
    Kr & $-12.60 \pm 0.07$ & $-581.62 \pm 0.00$ & $-9.63 \pm 0.11$ & $-1.71 \pm 0.13$ \\
    Xe & $-12.93 \pm 0.07$ & $-644.57 \pm 0.07$ & $-9.63 \pm 0.11$ & $-2.64 \pm 0.15$ \\
    \hline\hline
    \end{tabular}
    \end{table}

    The enthalpy components and corresponding formation enthalpies obtained from \textit{NPT} simulations at $T = 600$~K and $P = 500$~GPa are summarized in Table~\ref{tab:enthalpy_XH192}. While the alloyed systems exhibit negative total enthalpies, the relevant quantity for thermodynamic stability is the formation enthalpy $dH$, which displays a clear systematic trend. HeH$_{192}$ and NeH$_{192}$ show slightly positive $dH$ values within uncertainty, indicating marginal or unfavorable stability. In contrast, ArH$_{192}$, KrH$_{192}$, and XeH$_{192}$ exhibit increasingly negative formation enthalpies, demonstrating progressive stabilization with increasing noble-gas atomic number. This trend suggests stronger effective interactions between heavier noble gases and the dense hydrogen matrix under liquid metallic conditions.
    
    Importantly, this behavior contrasts sharply with the solid metallic hydrogen phase, where all noble-gas impurities exhibit positive formation free energies and are therefore thermodynamically unstable. The emergence of negative $dH$ values for Ar, Kr, and Xe in the liquid phase highlights the crucial role of liquid-state effects, particularly atomic disorder stabilization, in enabling the dissolution of heavier noble gases. In contrast, He and Ne are insoluble  both solid and liquid phases, indicating that liquid-phase stabilization is insufficient to overcome their unfavorable enthalpic contributions.
    
\subsection*{General Discussion}
The present results provide a microscopic framework for understanding noble-gas behavior in the deep interiors of giant planets. Our calculations demonstrate a pronounced phase dependence of noble-gas solubility in metallic hydrogen: all noble gases are thermodynamically unstable in the crystalline $I4_{1}/amd$ phase, whereas in the liquid phase heavier species (Ar, Kr, Xe) become stabilized while He and Ne remain insoluble. This is in contrast to molecular hydrogen: all noble gases mix with fluid hydrogen, and Ar,Kr,Xe also form solid hydrates at high pressure.  The general principle is that noble gases mix more easily in insulating molecular hydrogen than in metallic hydrogen.

In planetary interiors such as those of Jupiter and Saturn, metallic hydrogen is expected to exist predominantly as a dense fluid over a broad pressure range~\cite{guillot1999interiors,militzer2008massive,mcmahon2012properties}. The stabilization of heavier noble gases in liquid metallic hydrogen therefore suggests a mechanism by which Ar, Kr, and Xe may be retained within the metallic hydrogen envelope rather than segregating into distinct phases. Because these elements are chemically inert at ambient conditions, their stabilization arises not from conventional bonding but from collective electronic and entropic effects intrinsic to the dense quantum fluid. The competition between electronic repulsion, vibrational stiffening, and disorder-driven stabilization identified in this work provides a microscopic explanation for such behavior.

More broadly, the strong phase dependence uncovered in this study suggests that compositional stratification in giant planets may be sensitive not only to pressure and temperature, but also to the local phase state of hydrogen. In regions where metallic hydrogen transitions between crystalline and fluid states—whether due to thermal gradients, compositional effects, or dynamical processes—the solubility of trace species could change abruptly. Such variations may influence convective transport, density profiles, and long-term evolutionary pathways. 

The persistent demixing of He and Ne in fluid metallic hydrogen is particularly relevant for interpreting observed atmospheric anomalies. The depletion of neon in Jupiter’s atmosphere~\cite{mahaffy2000noble,owen1999low} has often been attributed to "helium-rain" processes and phase separation in hydrogen–helium mixtures~\cite{morales2009phase,lorenzen2009demixing}. Our results indicate that even in the absence of helium demixing, neon remains thermodynamically disfavored in metallic hydrogen, reinforcing the tendency for Ne to partition and, due to its higher mass, fall towards the core. The thermodynamically stable state could have a He-Ne core, with Ar, Kr, Xe soluble throughout the hydrogen region.  This is consistent with the observed Nwe deficiency and Ar,Kr,Xe excess in the Jovian atomosphere.

Finally, our findings emphasize that impurity thermodynamics in highly compressed quantum fluids cannot be inferred directly from ambient-pressure intuition. Even closed-shell species can become selectively stabilized through collective many-body effects in the liquid phase. By establishing a unified thermodynamic comparison between solid and liquid metallic hydrogen, this work provides a foundation for incorporating noble-gas solubility into future models of planetary interiors, core erosion, and atmospheric evolution.

\section*{Conclusions}
In this work, we have investigated the thermodynamic stability of noble-gas impurities in metallic hydrogen using \textit{ab initio} molecular dynamics and a first-principles free-energy framework. By explicitly comparing solid $I4_{1}/amd$ hydrogen with the corresponding liquid phase at 500 GPa, we demonstrate a strong phase dependence of noble-gas solubility under extreme conditions.

In the crystalline metallic phase, all noble gases considered (He, Ne, Ar, Kr, and Xe) exhibit positive formation free energies, indicating no solid solubility. This behavior is dominated by large electronic enthalpy penalties and reinforced by positive zero-point energy contributions, reflecting the absence of favorable bonding and the stiffening of  H$_2$ vibrational modes near the defect. In contrast, liquid metallic hydrogen displays selective stabilization of heavier noble gases. While He and Ne remain unstable, Ar, Kr, and Xe exhibit negative formation enthalpies, revealing that liquid-state disorder and entropic effects can overcome unfavorable electronic contributions for sufficiently large species.

These findings establish that noble-gas solubility in metallic hydrogen is intrinsically phase dependent. The selective retention of heavier noble gases in the liquid phase provides a plausible microscopic mechanism for noble-gas fractionation in giant-planet interiors: they are lifted from the core by their solubility in hydrogen.
The insolubility of He and Ne in metallic hydrogen is consistent with their observed depletion in the upper atmosphere: they will fall to the core region of the planet.  This means that "Neon rain" will exist, equivalent to "He-rain", although in either case rain is a misleading if poetic description of this  one-off process; precipitation would be a better term. 

More broadly, this work highlights the importance of explicitly accounting for phase state and thermodynamic contributions when modeling impurity behavior in dense quantum fluids under extreme compression.

\section*{Acknowledgments}
	This research project is supported by the Second Century Fund (C2F), Chulalongkorn University. We acknowledge the supporting computing infrastructure provided by NSTDA, CU, CUAASC, NSRF via PMUB [B05F650021, B37G660013] (Thailand). (\url{URL:www.e-science.in.th}). This also work used the ARCHER2 UK National Supercomputing Service (\url{https://www.archer2.ac.uk}) as part of the UKCP collaboration.

\bibliography{references}

\end{document}